\documentstyle[11pt]{article}\textheight 230mm\textwidth 150mm
            \newcommand{\be}{\begin{eqnarray}}
            \newcommand{\ee}{\end{eqnarray}}
            \newcommand{\eel}[1]{\label{#1}\end{eqnarray}}
\newcommand{\e}[1]{\label{e:#1}\end{eqnarray}}
     \newcommand{\eg}{{\em e.g.\ }}
            \newcommand{\ie}{{\em i.e.\ }}

            \newcommand{\la}{{\lambda}}
            
            \newcommand{\del}{{\delta}}

\newcommand{\cM}{{\cal{M}}}
\newcommand{\cU}{{\cal{U}}}

           \newcommand{\ra}{{\rightarrow}}
 \newcommand{\lea}{{\leftarrow}}
            
            \newcommand{\Lra}{{\Leftrightarrow}}
            \newcommand{\pet}{{\cal P}}

\newcommand{\ca}{{\cal C}}

            \newcommand{\beq}{\begin{quote}}
            \newcommand{\eq}{\end{quote}}
            \newcommand{\Om}{\Omega}

            \newcommand{\ben}{\begin{enumerate}}
            \newcommand{\een}{\end{enumerate}}
            \newcommand{\bit}{\begin{itemize}}
            \newcommand{\ei}{\end{itemize}}
    	\newcommand{\nn}{\nonumber}
            \newcommand{\r}[1]{(\ref{e:#1})}
            \newcommand{\edfl}[1]{\label{#1}\end{df}}
\def\theequation{\thesection.\arabic{equation}}

\newcommand{\ve}{{\varepsilon}}

\def\d{\partial}
\def\cC{{\cal C}}

 \def\cA{{\cal A}}
 \def\cS{{\cal S}}
 \def\cE{{\cal E}}

  \def\half{{1 \over 2}}

\begin{document}
\begin{titlepage}
\noindent
G\"{o}teborg ITP 98-08\\

\vspace*{5 mm}
\vspace*{35mm}
\begin{center}{\LARGE\bf Dualities between Poisson brackets and antibrackets}
\end{center} \vspace*{3 mm} \begin{center} \vspace*{3 mm}

\begin{center}Igor Batalin\footnote{On leave of absence from
P.N.Lebedev Physical Institute, 117924  Moscow, Russia\\E-mail:
batalin@td.lpi.ac.ru.} and Robert
Marnelius\footnote{E-mail: tferm@fy.chalmers.se}\\ \vspace*{7 mm}
{\sl Institute of Theoretical Physics\\
Chalmers University of Technology\\ G\"{o}teborg University\\
S-412 96  G\"{o}teborg, Sweden}\end{center}
\vspace*{25 mm}
\begin{abstract}
Recently it has been shown that antibrackets  may be expressed in
terms of  Poisson brackets and vice versa for commuting functions
in the original bracket. Here we also introduce generalized brackets
involving higher
antibrackets or higher Poisson brackets where the latter are of a new type.
We give
generating functions for these brackets for functions in arbitrary
involutions in the
original bracket. We also give
 master equations for generalized
Maurer-Cartan equations.  The presentation is completely symmetric with
respect to
 Poisson brackets and antibrackets.
\end{abstract}\end{center}\end{titlepage}

\setcounter{page}{1}
\setcounter{equation}{0}
\section{Introduction.}
Poisson brackets and antibrackets are two basic binary operations in classical
dynamics. Poisson brackets have an old history and are crucial for the
transition
to the quantum theory. Antibrackets \cite{anti,BV} on the other hand play a
crucial
role in the BV quantization of general gauge theories \cite{BV}. Recently
we have constructed a consistent quantum antibracket corresponding to the
commutator for Poisson brackets \cite{Quanti}. This construction did not only
cast new light on the BV quantization but also indirectly on the relations
between
antibrackets and Poisson brackets.  The purpose of the present paper
is to communicate these relations. The presentation will be completely symmetric
with respect to the two brackets and will therefore in a natural way also
include the
results in \cite{Ne,ND,AMS}. In this way we demonstrate the duality between
the two
brackets in the following sense: To any relation involving both brackets
there is a
dual relation in which the brackets are interchanged. As a result we present new
higher Poisson brackets constructed in complete analogy with higher
antibrackets.
Also the recently given quantum master equation for generalized Maurer-Cartan
equations given in \cite{OG} is shown to have a dual master equation at the
classical
level.

On a manifold, $\cE$, with dimension $(2n,2n)$ we may  have either a
nondegenerate Poisson bracket or a nondegenerate antibracket. (The defining
properties of these brackets are given in  appendices A and B.) If $\cE$ is a
symplectic manifold we have a Poisson bracket and the (local)
 Darboux coordinates with their Grassmann parities  ($\ve=0,1$ mod
2) may be defined to be
\be
&&z^A\equiv\{x^a, x^*_a, p_a, p^a_*\},\quad a=1,\ldots,n,\nn\\
&&\ve_a\equiv\ve(x^a)=\ve(p_a),\quad\ve(x^*_a)=\ve(p^a_*)=\ve_a+1.
\e{1}
In terms of these Darboux coordinates
  the Poisson bracket has the nonzero elements
\be
&&\{x^a, p_b\}=\del^a_b=-(-1)^{\ve_a}\{p_b, x^a\},\quad\{x^*_a,
p^b_*\}=\del^b_a=(-1)^{\ve_a}\{p^b_*, x^*_a\}.
\e{2}
Similarly, if $\cE$ is an
antisymplectic manifold we have an antibracket and the (local)
 Darboux coordinates \r{1} in terms of which
  the antibracket has the nonzero elements
\be
&&(x^a, x^*_b)=\del^a_b=-(x^*_b, x^a),
\quad (p_a, p^b_*)=\del^b_a=-(p^b_*, p_a).
\e{3}
One  basic question is whether or not one may have
 both a Poisson bracket and an antibracket on $\cE$. Our treatment here
will lead to
a negative answer to this question if the brackets are the conventional ones.
However, we will display some generalized possibilities of coexistence.

As an introductory remark let us for a moment assume that it is possible to
have both
brackets on
$\cE$. The question is then how they are supposed to be
related. In
\cite{Panti,KN} it was proposed that the equations of motions should be
possible to
write down both in terms of Poisson brackets and in terms of antibrackets. There
should then exist two Hamiltonians, one even ($H$) and one odd ($Q$) related by
\be
&&\{f(z), H\}=(f(z), Q).
\e{4}
However, this condition turns out to be very restrictive and allows only for
very special forms of $H$ and $Q$ \cite{Panti,KN}. Such relations are
therefore not
valid in general. On the other hand, it is amusing to note that if the two
brackets
for some reasons allow for the common Darboux coordinates \r{1} then there
exists a
very natural solution of
\r{4}, namely
\be
&&Q=p_ap^a_*(-1)^{\ve_a}-x^ax^*_a(-1)^{\ve_a},
\quad H=-x^ap_a-x^*_ap^a_*.
\e{5}
This solution satisfies the following supersymmetric Poisson bracket algebra and
antibracket algebra which seems to be a natural consequence of \r{4}
\cite{Panti,KN}.
\be
&&\{H,H\}=\{H, Q\}=0,\;\;\{Q, Q\}=2H,\nn\\&& (Q, Q)=
(H, Q)=0,\;\;(H, H)=2Q.
\e{6}
In addition we have  the relations
\be
&&(z^A, z^B)=\{z^A, \{Q, z^B\}\}=\half\left(\{z^A,
\{Q, z^B\}\}-\{z^B, \{Q,
z^A\}\}(-1)^{(\ve_A+1)(\ve_B+1)}\right),\nn\\
 &&\{z^A, z^B\}=(z^A, (H,
z^B))=\half\left((z^A, (H, z^B))-(z^B, (H, z^A))
(-1)^{\ve_A\ve_B}\right),
\e{7}
where $z^A$ are the Darboux coordinates \r{1}. However, the solution of the
coexistence problem is not that simple. In fact, there exists no relation
between
Poisson brackets and antibrackets  for {\em arbitrary} functions of $z^A$ of the
form \r{7} with $Q$ and $H$ given by \r{5}. On the other hand in
\cite{Quanti}-\cite{AMS} it has been demonstrated that there exist possible
ways to
express antibrackets in terms of Poisson brackets and vice versa but then for
{\em subclasses} of functions. Although this construction is similar to
\r{7} the
basic functions $Q$ and
$H$ do not satisfy
\r{6}. In this paper we present these possible ways to express antibrackets
in terms
of Poisson brackets and vice versa. In order to do this on the whole
manifolds  $\cE$ we are then naturally led to consider generalized brackets
defined in a definite way.

In  section 2 we  present the results in \cite{Quanti}-\cite{AMS} on how
ordinary
antibrackets may be expressed in terms of ordinary Poisson brackets and
vice versa.
This can only be done for the subclass of functions satisfying an abelian
algebra in
the original bracket. We also
give the precise circumstances under which
\r{4} is valid for arbitrary $H$ or
$Q$. In section 3 we present generating functions for Poisson brackets and
antibrackets for functions satisfying a Lie algebra in the original
brackets which
naturally lead us to generalized brackets involving higher antibrackets and
higher Poisson brackets which are valid on the whole manifold $\cE$.  In
section 4
generating functions of brackets of  functions in arbitrary involutions in the
original bracket are presented. We  demonstrate the existence of two dual master
equations for generalized Maurer-Cartan equations which naturally involve the
generalized brackets in section 3. In section 5 we then conclude the paper.
In two
appendices we give the defining properties of the conventional Poisson
brackets and
antibrackets.

\setcounter{equation}{0}
\section{Antibrackets in terms of Poisson brackets and vice versa.}
 Here we review
the results in \cite{Quanti}-\cite{AMS} in a unified manner which
illuminates the
possible ways to
express antibrackets in terms of Poisson brackets and vice versa. The
presentation is
in the spirit of \cite{Quanti}.

 Assume the existence of a graded Poisson
bracket {\em or} a graded antibracket on the
$(2n,2n)$-dimensional manifold
$\cE$. In terms of these nondegenerate brackets we define the following binary
operations
\be
&&(f, g)_Q\equiv\half\left(\{f,\{Q, g\}\}-\{g, \{Q,
f\}\}(-1)^{(\ve_f+1)(\ve_g+1)}\right),
\e{101}
\be
 &&\{f, g\}_R\equiv\half\left((f,(R, g))-(g, (R,
f))(-1)^{\ve_f\ve_g}\right),
\e{102}
where $Q$ is an odd function and $R$ an even function. These expressions satisfy
all the properties of a conventional antibracket and a conventional Poisson
bracket
except for the Jacobi identies and Leibniz' rule (see appendices A and B).
It is
easily seen that \eg Leibniz' rule require us to restrict the class of
functions.  Instead of Leibniz' rules
\r{a5} and
\r{a13} we have
\be
&&(fg, h)_Q-f(g, h)_Q-(f,
h)_Q\,g(-1)^{\ve_g(\ve_h+1)}=\nn\\
&&=\half\{f, h\}\{g,
Q\}(-1)^{\ve_h(\ve_g+1)}+\half\{f,Q\}\{g,h\}(-1)^{\ve_g},
\e{1021}
\be
&&\{fg, h\}_R-f\{g, h\}_R-\{f,
h\}_R\,g(-1)^{\ve_g\ve_h}=\nn\\
&&=-\half(f, h)(R,g)(-1)^{\ve_h(\ve_g+1)}-
\half(R,f)(g,h)(-1)^{\ve_g+\ve_h}.
\e{1022}
From these relations it follows that Leibniz' rule
is valid for
$(f, g)_Q$ if we restrict ourselves to the class of functions commuting in the
original Poisson bracket on
$\cE$ (denoted $\cM\subset\cE$ in the following), and that it is valid for $\{f,
g\}_R$ if we restrict ourselves to the class of functions commuting in the
original
antibracket on $\cE$ (denoted
$\cM^*\subset\cE\,$ in the following). On these submanifolds \r{101} and \r{102}
reduce to (cf.\cite{Quanti}-\cite{AMS} and \r{7})
\be
&&(f, g)_Q=\{f,\{Q, g\}\}\quad \forall f,g
\in\cM \quad(\Lra\; \{f,g\}=0),
\e{103}
\be
 &&\{f, g\}_R=(f,(R, g))\quad \forall f,g\in\cM^*
 \quad (\Lra\; (f,g)=0).
\e{104}
 In order for the expressions \r{101} and
\r{102} to also satisfy the  Jacobi identities we have to impose the
additional conditions
\cite{Quanti,AMS}
\be
&&\{\{f,\{g,\{ h, Q\}\}\}, Q\}-\half\{f,\{g,\{h, \{Q, Q\}\}\}\}=0,
\quad\forall f,g,h\in\cM,
\e{105}
\be
&&((f,(g,(h, R))), R)-\half(f,(g,(h, (R, R))))=0,
\quad \forall f,g,h\in\cM^*.
\e{106}
These conditions should be viewed as conditions on $Q$ and $R$. That there exist
solutions may easily be
 demonstrated in terms of  the Darboux
coordinates
\r{1} on $\cE$. We may let
$\cM$ be the class of functions which only depends on $x^a, x^*_a$.  In
this case
\r{103} is an antibracket on $\cM$ which coincide with the  antibracket \r{3} on
$\cM$ if $Q=p_ap^a_*(-1)^{\ve_a}$. After quantization this $Q$ is turned
into the
$\Delta$-operator in the BV-quantization
\cite{Quanti}. However, there are many other choices of $\cM$. For instance, we
may choose $\cM$ to be all functions of
$p_a, p^a_*$. In this case \r{103} is equal to the  antibracket \r{3} on this
submanifold $\cM$ if
$Q=-x^ax^*_a(-1)^{\ve_a}$. Similarly for \r{104} we may choose $\cM^*$ to be all
functions on $x^a, p_a$ in which case \r{104} is the  Poisson bracket \r{2} on
$\cM^*$ provided $R=-x^*_ap^a_*$, and if $\cM^*$ is all functions of $x^*_a,
p^a_*$ then \r{104} is equal to the  Poisson bracket \r{2} on this
$\cM^*$ provided $R=-x^ap_a$. All these choices of $Q$ and $R$ satisfy
\be
&&\{Q, Q\}=0,\quad (R, R)=0,
\e{107}
which are natural conditions on $Q$ and $R$.
 From now on we shall always require \r{107}.

The first conditions in
\r{105} and \r{106} are naturally restricted to the following conditions
\be
&&\{(f, g)_Q, h\}=0\;\;\Lra\;\;(f, g)_Q
\in\cM\quad\forall f,g,h\in\cM,
\e{1061}
\be
&&(\{f, g\}_R, h)=0\;\;\Lra\;\;\{f, g\}_R
\in\cM^*\quad\forall f,g,h\in\cM^*.
\e{1062}
These conditions make the brackets \r{103} and \r{104} closed on the
submanifolds
$\cM$ and
$\cM^*$ respectively.

That the formulas \r{101}-\r{102} or \r{103}-\r{104} may yield arbitrary
antibrackets and Poisson brackets may be seen as follows: Let $\cE$ be a
symplectic manifold with canonical coordinates $z^A, p_A$, $A=1,\ldots,2n$,
$\ve(z^A)=\ve(p_A)\equiv\ve_A$, satisfying
$\{z^A, p_B\}=\del^A_B$. Choose $\cM$ to
be all functions of $z^A$ and
\be
&&Q=-\half p_A E^{AB}(z) p_B (-1)^{\ve_B}.
\e{1071}
The formula \r{101} or \r{103} yields then the general antibracket \r{a15}. Note
that a nonzero odd $Q$ in \r{1071} requires the first equalities in \r{a16} and
that $\{Q, Q\}=0$ yields the Jacobi identities in \r{a16} since the first
condition in \r{105} is satisfied by \r{1071}. (The weaker condition in \r{105}
requires here
$\{Q, Q\}=0$.) The condition \r{1061} is valid for any $Q$ which is at most
quadratic in $p_A$. Similarly if
$\cE$ is an antisymplectic manifold with the
anticanonical coordinates $x^A, x^*_A$,
$A=1,\ldots,2n$,
$\ve(x^A)\equiv\ve_A$, $\ve(x^*_A)=\ve_A+1$, satisfying $(x^A, x^*_B)=\del^A_B$,
and if we choose $\cM^*$ to be all functions of $x^A$ and
\be
&&R=-\half x^*_A \omega^{AB}(x) x^*_B,
\e{1072}
then the formula \r{102} or \r{104}  yields the general Poisson bracket
\r{a7} on
$\cM^*$. A nonzero even $R$ requires the first properties in \r{a8} while $(R,
R)=0$ yields the Jacobi identities.  The
condition in \r{1062} is satisfied by any $R$ at most quadratic in $x^*_A$. This
connection to Poisson brackets was used in
\cite{BATU,ASZK}, and the formula \r{104}
was previously given in
\cite{Ne,ND,AMS}  and the corresponding relation in the $Sp(2)$ case was
given in
\cite{Trip}.

Thus, by means of a nondegenerate Poisson bracket/antibracket on $\cE$ we may
define a nondegenerate antibracket/Poisson bracket on $\cM$/$\cM^*$ provided
\r{105}/\r{106} is valid. The submanifolds $\cM$ and $\cM^*$ have half the size
of
$\cE$. The dimension of $\cM$ is $(n,n)$, and
the dimension of $\cM^*$ is $2n$ with
arbitrary distribution of fermions and bosons depending on the choice of $\cE$.
Note that the coexistence of an antibracket and a Poisson bracket is trivial
on these submanifolds: $\cM$ is an antisymplectic manifold on which the original
Poisson bracket is zero, while
$\cM^*$ is a symplectic manifold on which the original antibracket is zero.
Note that if we \eg calculate the Poisson
bracket \r{102} in terms of the antibracket
\r{101} then we get a Poisson bracket defined on a submanifold which is only
one-fourth of $\cE$. Furthermore, this Poisson bracket
is completely different from
the original one in \r{101}.

We end this section with a remark on equation \r{4} in the introduction.
In the above constructions of antibrackets
and Poisson brackets expressed in terms
of each other we may solve condition
\r{4} in the following sense: If we
 have an antibracket on $\cE$ and a Poisson
bracket on
$\cM^*$ there always exists an odd Hamiltonian
$Q_H$ on $\cE$ such that $\{f, H\}=(f,
Q_H)$ for all
$f,H\in\cM^*$  given by $Q_H=(R,H)$ (cf.\cite{Ne}).
 Example: Let $\cM^*$ be all
functions of
$x^a, p_a$ and $R=-x^*_ap^a_*$, then we have
\be
&&\{f(x,p), H(x,p)\}=(f(x,p), Q_H),
\quad Q_H=x^*_a\{x^a, H\}+p^a_*\{p_a, H\}.
\e{108}
Similarly, if we have a Poisson bracket
on $\cE$ and an antibracket on $\cM$ there
always exists a Hamiltonian H on $\cE$ for every odd
Hamiltonian $Q_H$ on $\cM$ given
by $H=\{Q, Q_H\}$. Example: Let $\cM$ be all functions of $x^a, x^*_a$  and
$Q=p_ap^a_*(-1)^{\ve_a}$, then we have
\be
&&(f(x,x^*), Q_H(x,x^*))=\{f(x,x^*), H\}, \nn\\
&& H=p_a(x^a,Q_H(x,x^*))+p^a_*(x^*_a,
Q_H(x,x^*)).
\e{109}
Thus,  we  always have counterparts to the Hamiltonians. However, these
counterparts are defined on an extended manifold.

\setcounter{equation}{0}
\section{Generating functions and  generalized  Poisson brackets and
antibrackets.}
The functions $Q$ and $R$ play a crucial role in the construction of an
antibracket on $\cM$ or a Poisson bracket on $\cM^*$ in terms of a nondegenerate
Poisson bracket or a nondegenerate antibracket on $\cE$ as presented in the
previous section.
$Q$ and
$R$ should satisfy
 $\{Q, Q\}=0$ and $(R, R)=0$. In addition they have to satisfy the
conditions \r{1061} and \r{1062}.  Here we propose a generalized scheme in
which
the latter conditions are relaxed. This scheme involves higher antibrackets and
higher Poisson brackets whose properties are most easily obtained by means of
generating functions of these brackets. (For quantum antibrackets these
relations
were given in \cite{Quanti,OG}.)

\subsection{Generating functions for generalized antibrackets.}
Let $\cE$ be a symplectic manifold. We have then a nondegenerate Poisson bracket
and an odd function $Q$ on $\cE$ satisfying $\{Q, Q\}=0$. Let us first
assume that
$f_a$ are functions on
$\cM$, \ie functions satisfying the abelian algebra $\{f_a, f_b\}=0$. We
may then
define the following canonically transformed
$Q$:
\be
&&Q(\phi)\equiv Q\exp{\{\stackrel{\lea}{ad}f_a \phi^a\}}, \quad
\ve(f_a)\equiv\ve_a=\ve(\phi^a),
\e{201}
where
\be
&&g \stackrel{\lea}{ad}f_a \equiv \{g, f_a\},
\e{202}
and where $\phi^a$ are parameters.
Obviously ($\d_a\equiv \d/\d\phi^a$)
\be
&& Q(0)=Q,\quad Q(\phi)\stackrel{\lea}{\d_a} = \{Q(\phi), f_a\}.
\e{204}
In terms of $Q(\phi)$ we may then define the
following generalized  antibrackets for
the functions on
$\cM$ (cf. \cite{hanti})
\be
&&(f_{a_1},
f_{a_2},\ldots,f_{a_n})_{Q(\phi)}\equiv-Q(\phi)
\stackrel{\lea}{\d}_{a_1}
\stackrel{\lea}{\d}_{a_2}\cdots
\stackrel{\lea}{\d}_{a_n}(-1)^{E_n}=
\nn\\&&=-\{\cdots\{\{Q(\phi),
f_{a_1}\},f_{a_2}\},\cdots, f_{a_n}\}
(-1)^{E_n},\quad E_n\equiv
\sum_{k=0}^{\left[{n-1\over 2}\right]}\ve_{a_{2k+1}}.
\e{205}
They satisfy  the properties
\be
&&(\ldots, f_a, f_b, \ldots)_{Q(\phi)}=
-(-1)^{(\ve_a+1)(\ve_b+1)}
(\ldots, f_b, f_a, \ldots)_{Q(\phi)}\nn\\
&&\ve((f_{a_1},
f_{a_2},\ldots,f_{a_n})_{Q(\phi)})=\ve_{a_1}+
\cdots+\ve_{a_n}+1.
\e{206}
Furthermore, since \r{201} is a canonically transformed $Q$ we have
\be
&&\{Q(\phi), Q(\phi)\}=\{Q, Q\}=0,
\e{207}
from which we may derive identities like
\be
&&\{Q(\phi), Q(\phi)\}\stackrel{\lea}{\d}_{a}
\stackrel{\lea}{\d}_{b}
\stackrel{\lea}{\d}_{c}(-1)^{\ve_b+
(\ve_a+1)(\ve_c+1)}\equiv 0,
\e{208}
which in turn implies
\be
&&(f_a, (f_b, f_c)_{Q(\phi)})_{Q(\phi)}
(-1)^{(\ve_a+1)(\ve_c+1)}+(f_c, (f_a,
f_b)_{Q(\phi)})_{Q(\phi)}
(-1)^{(\ve_c+1)(\ve_b+1)}+\nn\\&&+(f_b, (f_c,
f_a)_{Q(\phi)})_{Q(\phi)}(-1)^{(\ve_b+1)(\ve_a+1)}=
\{\{\{\{Q(\phi), f_a\}, f_b\},
f_c\}, Q(\phi)\}(-1)^{\ve_a\ve_c}\equiv\nn\\&&\equiv\{(f_a, f_b,
f_c)_{Q(\phi)}, Q(\phi)\}(-1)^{(\ve_a+1)(\ve_c+1)}.
\e{209}
It is easily seen that  condition
\r{1061} is equivalent to the vanishing of the 3-bracket ($n=3$) in
\r{205}. Thus,
in this case the 2-bracket in \r{205} is the conventional antibracket considered
in the previous section. Relation \r{209} is then just the  Jacobi
identities of the
2-antibracket.

Now if we relax
condition \r{1061} the formula \r{209} is not quite satisfactory as it
stands. The
reason is that the antibracket ($n=2$) in \r{209} is given by the expression
\r{103} which only is valid  for functions on $\cM$. However, if \r{1061} is
relaxed then $(f_a, f_b)_Q\notin\cM$ and the outer brackets in \r{209} are
{\em not}
the appropriate ones to use. Below we show that remarkably enough the
correct expression have the same form except for a factor minus one-half in
the last
equality.

A better definition of higher antibrackets are obtained if we consider  a more
general class of functions in the generating function
$Q(\phi)$. Let therefore the functions
$f_a$ in
\r{201} from now on be functions satisfying the Lie algebra
\be
&&\{f_a, f_b\}= U^c_{ab}f_c, \quad U^d_{ab}U^f_{dc}(-1)^{\ve_a\ve_c}+{
cycle}(a,b,c)\equiv 0.
\e{301}
where the structure coefficients $U^c_{ab}$ are constants. These functions
obviously
live on a larger manifold than $\cM$. Instead of \r{204} we have then
\be
&&Q(\phi)\stackrel{\lea}{\d}_a =\{Q(\phi),
\la^b_a(\phi)f_b(-1)^{\ve_a+\ve_b}\},
\e{302}
where integrability requires the parameter
matrices $\la^b_a(\phi)$ to satisfy the
Maurer-Cartan equation
\be
&&\d_a\la_b^c-\d_b\la_a^c(-1)^{\ve_a\ve_b}=\la^e_a\la^d_b
U^c_{de}(-1)^{\ve_b\ve_e+\ve_c+\ve_d+\ve_e},\quad \la^b_a(0)=\del^b_a.
\e{303}
The explicit solution is
\be
&&\tilde{\la}=(1-e^{-X})/X, \quad
\tilde{\la}_b^a\equiv\la_b^a(-1)^{\ve_a(\ve_b+1)},
\;\; X_b^a\equiv\phi^cU^a_{cb}(-1)^{\ve_c+\ve_b(\ve_a+1)},
\e{304}
where
$\phi^b\la_b^a=\phi^a$. Generalized higher
antibrackets are then obtained from the
definition
\r{205}. In particular we have
\be
&&(f_a,
f_b)'_{Q(\phi)}\equiv-Q(\phi)\stackrel{\lea}{\d}_a
\stackrel{\lea}{\d}_b(-1)^{\ve_a}=\nn\\
&&=-\la^c_a\la^d_b\{\{Q(\phi), f_d\},
f_c\}(-1)^{\ve_b\ve_c+\ve_a}-(\d_b\la^c_a)\{Q(\phi),
f_c\}(-1)^{\ve_a(\ve_b+1)}.
\e{305}
This
expression reduces exactly to the general ansatz \r{101} at $\phi^a=0$. However,
compared to the case in section 2 we have now a complete control over the Jacobi
identities since
\r{207}   is valid leading to identities like \r{208} involving the true
violation of the Jacobi identities when \r{1061} is relaxed.
Eq.\r{208} yields now at $\phi^a=0$
\be
&&(f_a, (f_b, f_c)_{Q})_{Q}(-1)^{(\ve_a+1)(\ve_c+1)}+(f_c, (f_a,
f_b)_{Q})_{Q}(-1)^{(\ve_c+1)(\ve_b+1)}+\nn\\&&+(f_b, (f_c,
f_a)_{Q})_{Q}(-1)^{(\ve_b+1)(\ve_a+1)}=-\half\{(f_a, f_b,
f_c)_{Q}, Q\}(-1)^{(\ve_a+1)(\ve_c+1)},
\e{3051}
where the higher
antibracket is given by
\be
&&(f_a, f_b,
f_c)_{Q}(-1)^{(\ve_a+1)(\ve_c+1)}\equiv\left.(f_a, f_b,
f_c)'_{Q(\phi)}(-1)^{(\ve_a+1)(\ve_c+1)}
\right|_{\phi=0}=\nn\\&&={1\over3}\left(\{(f_a,
f_b)_Q, f_c\}(-1)^{\ve_c+(\ve_a+1)(\ve_c+1)}+cycle(a,b,c)\right),
\e{3052}
where in turn the
antibrackets are given by
\r{101}. Although this 3-antibracket is defined as in \r{205} it differs
from the
explicit expression on the right-hand side of \r{205} since all
antibrackets now are
given by
\r{101}. In fact,  the definition \r{3052} as well as the  relation \r{3051} are
valid for arbitrary dynamical functions, \ie not just of the class \r{301}.
(Still
higher generalized antibrackets may also be expressed in terms of lower
ones.) It is
remarkable that
\r{3051} differs from \r{209} only by a factor.

 Note that
the class of functions satisfying \r{301} also includes all products of such
functions: Define
$F_A$  to be all monomials of $f_a$,
\ie
$F_A\equiv f_a, f_af_b, f_af_bf_c,\ldots$. These operators do also satisfy a
nonabelian Lie algebra,
$\{F_A, F_B\}=\cU^C_{AB}F_C$.
In accordance with \r{201} we may therefore define
\be
&&Q(\Phi)\equiv
Q\,\exp{\{\stackrel{\lea}{ad}F_A\Phi^A\}},
\e{306}
where the parameters $\Phi^A$ are $\phi^a,
\phi^{ab}, \phi^{abc}, \ldots$. Since
$F_A$ satisfies a nonabelian Lie
algebra we have also here integrable equations of the
form
\r{302} and generalized  antibrackets (${\d}_A=\d/\d\Phi^A$)
\be
&&(F_A, F_B)'_{Q(\Phi)}\equiv
-Q(\Phi)\stackrel{\lea}{\d}_A
\stackrel{\lea}{\d}_B(-1)^{\ve_A}, \quad
\ve_A\equiv
\ve(F_A)=\ve(\Phi^A),
\e{307}
which again reduces to \r{101} at $\Phi^A=0$.

 Although we have found a consistent generalized scheme in terms of higher
antibrackets it should be noted that these generalized antibrackets do not in
general satisfy Leibniz' rule
\r{a13}. The violation is given by \r{1021}. For the class of functions
satisfying
the Lie algebra \r{301} we have in particular
\be
&&(f_af_b, f_c)_Q-f_a(f_b, f_c)_Q-(f_a,
f_c)_Qf_b(-1)^{\ve_b(\ve_c+1)}=\nn\\ &&=\half\left(\{f_a, Q\}
 U_{bc}^d (-1)^{\ve_b} +
 \{f_b, Q\}U_{ac}^d(-1)^{\ve_a(\ve_b+1)}\right)f_d,
\e{308}
which in the case when $f_a$ are interpreted as constraint variables may be
viewed as
a weak violation since the right-hand side vanishes on the constraint surface,
$f_a=0$. One may notice that if a class of functions satisfy Leibniz' rule
for the
2-bracket  it is
 satisfied for all higher antibrackets as well.

\subsection{Generating functions for generalized Poisson brackets.}
There exists a completely dual construction when $\cE$ is an antisymplectic
manifold. In this case we have a nondegenerate antibracket
and an even function $R$ on $\cE$ satisfying $(R, R)=0$.
 Let to start with $f_a$ be
functions on
$\cM^*$ satisfying $(f_a, f_b)=0$. We may
then define the following anticanonically
transformed
$R$:
\be
&&R(\phi)\equiv R\exp{\{\stackrel{\lea}{Ad}f_a \phi^a\}},
 \quad \ve(f_a)\equiv\ve_a,
\quad\ve(\phi^a)=\ve_a+1,
\e{210}
where
\be
&&g \stackrel{\lea}{Ad}f_a \equiv (g, f_a),
\e{211}
and where $\phi^a$ are parameters.
Obviously
\be
&& R(0)=R,\quad R(\phi)\stackrel{\lea}{\d_a} = (R(\phi), f_a).
\e{213}
In terms of $R(\phi)$ we define the following higher Poisson brackets for
functions
on $\cM^*$
\be
&&\{f_{a_1},
f_{a_2},\ldots,f_{a_n}\}_{R(\phi)}\equiv -R(\phi)
\stackrel{\lea}{\d}_{a_1}\stackrel{\lea}{\d}_{a_2}\cdots
\stackrel{\lea}{\d}_{a_n}(-1)^{E_n}=\nn\\&&=-(\cdots((R(\phi),
f_{a_1}),f_{a_2}),\cdots, f_{a_n})(-1)^{E_n},\quad E_n\equiv
\sum_{k=0}^{\left[{n-1\over 2}\right]}(\ve_{a_{2k+1}}+1).
\e{214}
 Note the
properties
\be
&&\{\ldots, f_a, f_b, \ldots\}_{R(\phi)}=-(-1)^{\ve_a\ve_b}
\{\ldots, f_b, f_a, \ldots\}_{R(\phi)},\nn\\
&&\ve(\{f_{a_1},
f_{a_2},\ldots,f_{a_n}\}_{R(\phi)})=\ve_{a_1}+\cdots+\ve_{a_n}+n.
\e{215}
To our knowledge these higher Poisson brackets are of a new kind. Nambu once
introduced similar higher  Poisson brackets \cite{Nambu}
but they were not defined as in
\r{214}.

 Since
\r{210} is an anticanonically transformed
$R$ we have
\be
&&(R(\phi), R(\phi))=(R, R)=0,
\e{216}
from which we may derive identities like
\be
&&(R(\phi), R(\phi))\stackrel{\lea}{\d}_{a}\stackrel{\lea}{\d}_{b}
\stackrel{\lea}{\d}_{c}(-1)^{\ve_b+\ve_a\ve_c}\equiv 0,
\e{217}
which in turns implies
\be
&&\{f_a, \{f_b, f_c\}_{R(\phi)}\}_{R(\phi)}(-1)^{\ve_a\ve_c}+\{f_c, \{f_a,
f_b\}_{R(\phi)}\}_{R(\phi)}(-1)^{\ve_c\ve_b}+\nn\\&&+\{f_b, \{f_c,
f_a\}_{R(\phi)}\}_{R(\phi)}(-1)^{\ve_b\ve_a}=((((R(\phi), f_a), f_b),
f_c), R(\phi))(-1)^{(\ve_a+1)(\ve_c+1)}\equiv\nn\\&&\equiv(\{f_a, f_b,
f_c\}_{R(\phi)}, R(\phi))(-1)^{\ve_a\ve_c}.
\e{218}
It is easily seen that  condition
\r{1062} is equivalent to the vanishing of the 3-bracket ($n=3$) in
\r{214}. Thus,
in this case the 2-bracket in \r{214} is the conventional Poisson bracket
considered
in the previous section. Relation \r{218} is then just the  Jacobi
identities of the
 Poisson bracket.

Similarly to the case in the previous section  formula \r{218} is
 not quite satisfactory  if we relax
condition \r{1062} since the Poisson bracket ($n=2$) in \r{218} is given by
\r{104} which is valid only for functions belonging to $\cM^*$. However, if
\r{1062}
is relaxed then $(f_a, f_b)_Q\notin\cM$ and the outer brackets in \r{218}
are {\em
not} the appropriate ones. Below it it is shown that the correct expression
have the
same form except for a factor one-half in the last equality.

Replace the functions in the generating function $R(\phi)$ by  functions $f_a$,
$\ve(f_a)=\ve_a$,
 satisfying the antibracket Lie algebra
\be
&&(f_a, f_b)= U^c_{ab}f_c, \quad U^d_{ab}
U^f_{dc}(-1)^{(\ve_a+1)(\ve_c+1)}+{
cycle}(a,b,c)\equiv 0.
\e{309}
Note that the structure coefficients here are different from those in
\r{301} due
to the oddness of the antibracket
($\ve(U^c_{ab})=\ve_a+\ve_b+\ve_c+1$).
Instead of \r{213} we have
\be
&&R(\phi)\stackrel{\lea}{\d}_a =(R(\phi),
 \la^b_a(\phi)f_b(-1)^{\ve_a+\ve_b}),\quad
\ve(\la^b_a)=\ve_a+\ve_b,
\e{310}
where integrability requires the parameter matrices $\la^b_a(\phi)$ ($
\la^b_a(0)=\del^b_a$) to satisfy
the following Maurer-Cartan equation which is of a new type (cf.\r{303})
\be
&&\d_a\la_b^c+\d_b\la_a^c(-1)^{\ve_a\ve_b}=(-1)^{\ve_a}\la^e_a\la^d_b
(-1)^{(\ve_b+1)(\ve_e+1)}U^c_{de}(-1)^{\ve_c+\ve_d+\ve_e}.
\e{311}
The explicit solution is
\be
&&\tilde{\la}=(1-e^{-X})/X, \quad
\tilde{\la}_b^a\equiv\la_b^a(-1)^{\ve_a(\ve_b+1)},
\;\; X_b^a\equiv\phi^cU^a_{cb}(-1)^{\ve_b(\ve_a+1)},
\e{312}
where now
$\phi^b\la_b^a(-1)^{\ve_b}=\phi^a(-1)^{\ve_a}$. Note that we
have here $\ve(\phi^a)=\ve_a+1$. Generalized higher Poisson
brackets are defined by
\r{214}. In particular we have therefore
\be
&&\{f_a,
f_b\}'_{R(\phi)}\equiv R(\phi)\stackrel{\lea}{\d}_a
\stackrel{\lea}{\d}_b(-1)^{\ve_a}=\nn\\
&&=\la^c_a\la^d_b((R(\phi), f_d),
f_c)(-1)^{(\ve_b+1)(\ve_c+1)+\ve_a}-(\d_b\la^c_a)(R(\phi),
f_c)(-1)^{\ve_a\ve_b}.
\e{313}
This
expression reduces exactly to the general ansatz \r{102} at $\phi^a=0$, and
again we have  a complete control over the Jacobi
identities since
\r{216}   is valid leading to identities
like \r{217} now involving the violation of
the true Jacobi identities. We have explicitly
\be
&&\{f_a, \{f_b, f_c\}_{R}\}_{R}(-1)^{\ve_a\ve_c}+\{f_c, \{f_a,
f_b\}_{R}\}_{R}(-1)^{\ve_c\ve_b}+\nn\\&&+\{f_b, \{f_c,
f_a\}_{R}\}_{R}(-1)^{\ve_b\ve_a}=\half(\{f_a, f_b,
f_c\}_{R}, R)(-1)^{\ve_a\ve_c},
\e{3131}
where the
higher Poisson bracket is given by
\be
&&\{f_a, f_b,
f_c\}_{R}(-1)^{\ve_a\ve_c}\equiv\left.\{f_a, f_b,
f_c\}'_{R(\phi)}(-1)^{\ve_a\ve_c}
\right|_{\phi=0}=\nn\\&&={1\over3}\left((\{f_a,
f_b\}_R, f_c)(-1)^{\ve_c+\ve_a\ve_c}+cycle(a,b,c)\right),
\e{3132}
where in turn the Poisson brackets are given by \r{102}.  Note that
\r{3131} differs from
\r{209} only by a factor.
Also here the class of
functions satisfying
\r{309}  includes all products of such functions.

Instead of Leibniz' rule \r{a5} we have from
\r{1022}
\be
&&\{f_af_b, f_c\}_R-f_a\{f_b, f_c\}_R-\{f_a,
f_c\}_Rf_b(-1)^{\ve_b\ve_c}=\nn\\ &&=
-\half\left(
(R, f_a)
 U_{bc}^d (-1)^{\ve_b+\ve_c} +
 (R, f_b)U_{ac}^d(-1)^{(\ve_a+1)(\ve_b+1)}\right)f_d,
\e{314}
which in the case when $f_a$ are interpreted as constraint variables may be
viewed as
a weak violation of Leibniz' rule since it vanishes on the constraint surface,
$f_a=0$. One may notice that if a class of functions satisfy Leibniz' rule
for the
Poisson bracket  it is
 satisfied for all higher Poisson brackets as well.

\subsection{The most general brackets}
If we  consider the brackets \r{101} and \r{102} on the whole of the original
manifold $\cE$ and require $Q$ and $R$ to satisfy \r{107} then we have a
consistent
scheme of generalized brackets involving a tower of higher brackets on
$\cE$. These
higher brackets are defined by \r{3052} etc and \r{3132} etc which may be
extracted
from the generating functions of the higher brackets for functions
satisfying Lie
algebra relations in the original bracket. The Jacobi identities of \r{101} and
\r{102} are violated by the expressions \r{3051} with \r{3052} and \r{3131} with
\r{3132}, which remarkably enough are valid for arbitrary functions on
$\cE$. The
corresponding relations for the higher brackets of functions satisfying Lie
algebra
relations, which are obtained by identities like \r{208} and \r{217},
should  also be
valid for arbitrary functions. However, note that the brackets
\r{101} and \r{102} violate Leibniz' rule by the expressions \r{1021} and
\r{1022}.

\setcounter{equation}{0}
\section{Generalized Maurer-Cartan equations}
By means of generating functions of
brackets of functions satisfying a Lie algebra in
the original brackets we were led to
the most general definition of generalized
antibrackets and Poisson brackets in
terms of higher brackets at the end of the last
section. Although there are no
generating functions for such brackets involving
arbitrary functions on $\cE$ there is
a further generalization of generating
functions which generate the above
 brackets for functions in arbitrary involutions in
the original bracket. This generalization
involves a lot of interesting features like
 a new
type of master equation encoding new
generalized Maurer-Cartan equations. This is
presented below. (For quantum
antibrackets this construction was given in \cite{OG}.)

\subsection{Functions in arbitrary involutions in a Poisson bracket sense.}
 Let
$\cE$ be a symplectic manifold and let
$f_a$ be functions on
$\cE$ satisfying the Poisson algebra
\be
&&\{f_a, f_b\}= U^c_{ab}f_c,\quad \ve(f_a)\equiv\ve_a,
\e{401}
where the structure coefficients
$U^c_{ab}$ now may be arbitrary dynamical
functions. This case is most
efficiently treated if we extend the original phase
space $\cE$ by the ghost variables $\cC^a, \pet_a$,
$\ve(\cC^a)=\ve(\pet_a)=\ve_a+1$, satisfying
$\{\cC^a,
\pet_b\}=\del^a_b$. We may then always
define an odd nilpotent function $\Om$ with
ghost number one according to the prescription
\be
&&\Om=\cC^a f_a+\ldots, \quad \{\Om, \Om\}=0,\nn\\
&&\{G,\Om\}=\Om,\quad G\equiv \cC^a\pet_a(-1)^{\ve_a},
\e{402}
where $G$ is the ghost charge and where the additional terms in
$\Om$ depends on
$\pet_a$ (see
\cite{BFV}). ($\Om$ is the BFV-BRST charge.)

Instead of the explicit form \r{201}
of the generating function $Q(\phi)$ used in the
previous sections, we define here
$Q(\phi)$ by the differential equation
\be
&&Q(\phi)\stackrel{\lea}{\d_a}=
\{Q(\phi), Y_a(\phi)\}, \quad Q(0)\equiv Q,
\e{403}
which is similar to \r{302}.
The connections $Y_a(\phi)$,
$\ve(Y_a)=\ve_a$, satisfy the integrability
conditions (zero curvature condition)
\be
&&Y_a\stackrel{\lea}{\d_b}-Y_b
\stackrel{\lea}{\d_a}(-1)^{\ve_a\ve_b}=\{Y_a, Y_b\},
\e{404}
which in turn are integrable.
Note that \r{403} implies
\be
&&\{Q(\phi), Q(\phi)\}
\stackrel{\lea}{\d_a}=-\{Y_a(\phi), \{Q(\phi), Q(\phi)\}\}.
\e{405}
Thus, the boundary condition
$\{Q(0), Q(0)\}=0$ implies $\{Q(\phi), Q(\phi)\}=0$,
which shows that $Q\,\ra\,
Q(\phi)$ is a canonical transformation.

Generalized antibrackets may be defined by
\r{205} with
$f_a$ replaced by
$Y_a$. In particular we have
\be
&&(Y_a(\phi),
Y_b(\phi))'_{Q(\phi)}\equiv-Q(\phi)\stackrel{\lea}{\d_a}
\stackrel{\lea}{\d_b}(-1)^{\ve_a}=\nn\\
&&=\half\left(\{Y_a,
\{Q(\phi), Y_b\}\}-\{Y_b, \{ Q(\phi),
Y_a\}(-1)^{(\ve_a+1)(\ve_b+1)}\right)-\nn\\
&&-\half\{Q(\phi),
Y_a\stackrel{\lea}{\d_b}+Y_b
\stackrel{\lea}{\d_a}(-1)^{\ve_a\ve_b}\}(-1)^{\ve_a},
\e{406}
which should coincide with \r{101} at
 $\phi^a=0$.  We expect therefore
the last line in \r{406} to vanish at
$\phi^a=0$.  Indeed, this is the case
 for the first rank quasigroup theories
 treated below.  For the next higher
antibracket we find
\be
&&(Y_a(\phi), Y_b(\phi), Y_c(\phi))'_{Q(\phi)}
 (-1)^{(\ve_a + 1)(\ve_c + 1)} \equiv
 Q(\phi) \stackrel{\lea}{\d_a}
\stackrel{\lea}{\d_b} \stackrel{\lea}{\d_c}
 (-1)^{\ve_a
\ve_c} =\nn\\&& = {1\over 3}
\left(\{(Y_a, Y_b)'_Q, Y_c\} (-1)^{\ve_c + (\ve_a +
1)(\ve_c + 1)} + cycle (a, b, c)\right) +
\nn\\&&+ {1\over 3} \left(\{\{Q,
Y_a\}, Y_b
\stackrel{\lea}{\d_c} + Y_c
\stackrel{\lea}{\d_b} (-1)^ {\ve_b \ve_c}\} (-1)^{\ve_a
\ve_c} + cycle (a, b, c)\right) +\nn\\&&
+ {1\over 6} \left(\{Q, (Y_a
\stackrel{\lea}{\d_b} + Y_b
\stackrel{\lea}{\d_a} (-1)^{\ve_a
\ve_b})
\stackrel{\lea}{\d_c}\} (-1)^{\ve_a
\ve_c} + cycle (a, b, c)\right),
\e{407}
which differ from \r{3131} by similar terms as
\r{406} differ from \r{101}. Again we
expect these deviations to vanish at $\phi^a=0$.
The same situation is expected to
occur for all higher antibrackets, \ie they
 are expected to coincide with the
generalized higher antibrackets of section 4 at $\phi^a=0$.
As we already have mentioned in the previous
 section the generalized  brackets above
do not satisfy Leibniz' rule. However,
for the functions $Y_a$  we have from \r{1021}
\be
&&(Y_aY_b, Y_c)_Q-Y_a(Y_b, Y_c)_Q-(Y_a,
Y_c)_QY_b(-1)^{\ve_b(\ve_c+1)}=\nn\\ &&=\half\{Y_a, Q\}
 \{Y_b, Y_c\}(-1)^{\ve_b} +
 \half\{Y_a, Y_c\}\{Y_b, Q\}(-1)^{\ve_c(\ve_b+1)},
\e{4071}
where the right-hand side vanishes on the
hypersurface determined  by $\pet_a=0$ and
$\theta_a=0$.

Following \cite{OG} we propose now the following
manifestly BRST invariant expression
for the connections $Y_a$ in
\r{403}
\be
&&Y_a(\phi)\equiv\{\Om, \Om_a(\phi)\},
\quad \ve(\Om_a)=\ve_a+1.
\e{408}
The integrability conditions \r{404}
 become then in terms of $\Om_a$
\be
&&\Om_a\stackrel{\lea}{\d_b}-\Om_b
\stackrel{\lea}{\d_a}(-1)^{\ve_a\ve_b}=(\Om_a,
\Om_b)_{\Om}-\half\{\Om_{ab}, \Om\},
\e{409}
where we have introduced the $\Om$-antibracket
defined in accordance with \r{101},
\ie
\be
&&(A, B)_{\Om}\equiv\half
\left(\{A, \{\Om, B\}\}-\{B, \{\Om,
A\}\}(-1)^{(\ve_A+1)(\ve_B+1)}\right).
\e{410}
We expect
$Y_a(\phi)$ to be of the form
\be
&&Y_a(\phi)=\la^b_a(\phi)f_b(-1)^{\ve_a+\ve_b}+
\{\mbox{\small possible ghost
dependent terms}\},\; \la^b_a(0)=\del^b_a,
\e{411}
which makes \r{403} similar to \r{302}, although
$\la^b_a(\phi)$ may be dynamical
functions here. The definition
\r{408} requires then $\Om_a$ to be of the form
\be
&&\Om_a(\phi)=\la^b_a(\phi)\pet_b+\{\mbox{\small
possible ghost dependent terms}\}.
\e{412}
This means in turn that \r{409} are
generalized Maurer-Cartan equations for
$\la^b_a(\phi)$, $\ve(\la^b_a)=\ve_a+\ve_b$.
 The integrability conditions of \r{409}
involve first derivatives of $\Om_{ab}$,
 a third order antibracket, and a new
function $\Om_{abc}$. The subsequent
integrability conditions involve
 still higher
$\Om$-antibrackets and functions
$\Om_{abc\ldots}$ with still more indices. However,
these integrability conditions are only
implicit equations for the functions
$\Om_{abc\ldots}$.

Following the treatment in \cite{OG}
we propose below in \r{416} one single master
equation involving only two basic
 dynamical functions which determines
$\Om_a$,
$\Om_{ab}$, etc. One of these dynamical
 functions is an extended BFV-BRST charge
$\Delta$ defined by
\be
&&\Delta\equiv\Om+\eta^a \pi_a(-1)^{\ve_a},
\quad \{\Delta, \Delta\}=0,\nn\\
&&\{\phi^a, \pi_b\}=\del^a_b, \quad \ve(\eta^a)=
\ve_a+1,\quad \ve(\pi_a)=\ve_a,
\e{413}
where we have introduced the canonical
conjugate variables $\pi_a$
to $\phi^a$, now turned dynamical,
 and the new ghost variables $\eta^a$,
 to be treated as parameters.
The other basic function is an
even, extended ghost charge defined by
\be
&&S(\phi, \eta)\equiv G+\eta^a\Om_a(\phi)+\half
\eta^b\eta^a\Om_{ab}(\phi)(-1)^{\ve_b}+
\nn\\&&+{1\over6}\eta^c\eta^b\eta^a\Om_{abc}
(\phi)(-1)^{\ve_b+\ve_a\ve_c}
+\cdots\nn\\&&
\cdots+
{1\over n!}\eta^{a_n}\cdots\eta^{a_1}\Om_{a_1\cdots
a_n}(\phi)(-1)^{(\ve_{a_2}+\ldots+\ve_{a_{n-1}}
+\ve_{a_1}\ve_{a_n})}+\cdots,
\e{414}
where $G$ is the ghost charge in \r{402}.
$\Om_{a_1\cdots a_n}(\phi)$ are
dynamical functions with the properties
\be
&&\ve(\Om_{a_1\cdots a_n}(\phi))=\ve_{a_1}+
\cdots+\ve_{a_n}+n, \quad \{G,
\Om_{a_1\cdots a_n}\}=-n\Om_{a_1\cdots a_n}.
\e{415}
Thus, $\Om_{a_1\cdots a_n}$ has  ghost number
$-n$, which means that if we assign ghost
number one to $\eta^a$ then $\Delta$
 has ghost number one and $S$ has ghost number
zero. We propose now that the functions
$\Om_{a_1\cdots a_n}(\phi)$ are
related to the ones in the integrability conditions
above and that they are determined
by the master equation
\be
&&(S, S)_{\Delta}=\{\Delta, S\},
\e{416}
where the antibracket is
defined by \r{101}, \ie we have
\be
&&(S, S)_{\Delta}\equiv\{\{ S, \Delta\}, S\}.
\e{417}
Consistency requires  $\{\Delta, S\}$ to
 be nilpotent in a Poisson bracket
sense since
\be
&&\{\Delta, (S, S)_{\Delta}\}=0\quad
\Lra\quad \{\{\Delta, S\}, \{\Delta, S\}\}=0.
\e{418}
The explicit form of $\{S, \Delta\} (=-\{\Delta, S\})$ is
\be
&&\{S, \Delta\}=\Om+\eta^a\{\Om_a,
\Om\}+\eta^b\eta^a\Om_a\stackrel{\lea}{\d_b}(-1)^{\ve_b}+
\half\eta^b\eta^a\{\Om_{ab},
\Om\}(-1)^{\ve_b}+\nn\\&&+\half\eta^c\eta^b\eta^a\Om_{ab}
\stackrel{\lea}{\d_c}(-1)^{\ve_b+\ve_c}+{1\over
6}\eta^c\eta^b\eta^a\{\Om_{abc},
\Om\}(-1)^{\ve_b+\ve_a\ve_c}+O(\eta^4).
\e{419}
Note that \r{416} may be written as
$\{S, \{S, \Delta\}\}=\{S, \Delta\}$ which when
compared with $\{G, \Omega\}=\Omega$ is
consistent with $S$ as an extended ghost
charge and with $\{S, \Delta\}$ as an extended BRST charge.

 To zeroth and first order
in
$\eta^a$ the master equation \r{416} is
satisfied identically. However, to second
order in $\eta^a$ it yields exactly the integrability conditions \r{409}.
 At the third order in $\eta^a$ it yields
\be
&&\Om_{bc}\stackrel{\lea}{\d_a}(-1)^{(\ve_b+1)\ve_a}+\half(\Om_a,
\Om_{bc})_{\Om}(-1)^{\ve_a\ve_c}+cycle(a,b,c)=\nn\\&&=-(\Om_a,
\Om_b,
\Om_c)_{\Om}(-1)^{\ve_a\ve_c}-{2\over3}\{\Om'_{abc},\Om\},\nn\\
&&
\Om'_{abc}\equiv\Om_{abc}-{1\over8}\left(\{\Om_{ab},
\Om_c\}(-1)^{\ve_a\ve_c}+cycle(a,b,c)\right),
\e{420}
where
the higher  $\Om$-antibracket is defined
in accordance with \r{3051}. The above
equations are indeed consistent with the
integrability conditions of \r{409}. At
higher orders in
$\eta^a$ the master equation
\r{416} yields equations involving still higher
$\Om$-antibrackets and functions
$\Om_{abc\ldots}$ with still more indices.
 These equations we
expect to be consistent with the integrability conditions of
\r{420}. This chain of
integrability conditions will then terminate as soon as a
higher order antibracket is zero.
 We expect this level to be  connected to the rank
of the theory, \ie the highest power of
$\pet_a$ in $\Om$. For instance, if $\Om$ is
of rank
$r$  it would be natural to have $(\Om_{a_1}, \Om_{a_2},
\ldots, \Om_{a_n})_{\Om}=0$ for $n>r+1$ in
which case $\Om_{a_1a_2a_3\cdots a_n}=0$
for
$n>r$ (see below).

As an illustration of our formulas we
consider now functions $f_a$ forming a
quasigroup rank one theory. In this case we have (cf.\cite{OG})
\be
&&\Om=\cC^a f_a+\half \cC^b\cC^a
U^c_{ab}\pet_c(-1)^{\ve_c+\ve_b}.
\e{421}
where $\{\Om, \Om\}=0$ requires 
$\{f_a, f_b\}= U^c_{ab}f_c$ and
\be
&&\left(U^d_{ab}U^e_{dc}+\{U^e_{ab},
f_c\}(-1)^{\ve_c\ve_e}\right)
(-1)^{\ve_a\ve_c}+{cycle}(a,b,c)\equiv 0
\e{422}
plus conditions on $\{U^c_{ab}, U^f_{de}\}$.
 The latter conditions are satisfied if
\be
&&\{U^c_{ab}, U^f_{de}\}=0, \quad \{\{f_d,
U^c_{ab}\}, U^g_{ef}\}=0,
\e{423}
which are stronger conditions
than what is required by $\{\Om, \Om\}=0$.
Here $\Om_a$ may be chosen to be
\be
&&\Om_a(\phi)=\la^b_a(\phi)\pet_b,
\quad \la^b_a(0)=\del^b_a,
\e{424}
where we assume that
\be
&&\{\la^b_a(\phi), \la^d_c(\phi)\}=0,
\quad \Lra \quad \{\Om_a(\phi),
\Om_b(\phi)\}=0.
\e{425}
We require now
\be
&&(\Om_a, \Om_b, \Om_c)_{\Om}=0\quad
\Lra \quad \{(\Om_a, \Om_b)_{\Om}, \Om_c\}=0,
\e{426}
which allows for the choice $\Om_{ab}=0$ since \r{426} makes the antibracket
$(\Om_a, \Om_b)_{\Om}$ satisfy the Jacobi identities, which in turn makes
\r{409}
integrable without the need of a nontrivial $\Om_{ab}$. All higher
integrability conditions are then identically zero. Condition
\r{426} is satisfied if we impose
\be
&&\{\la^b_a(\phi),
U^c_{de}\}=0,
\quad \{\la^b_a,\{\la^d_c,
f_e\}\}=0.
\e{427}
All the above conditions specify  quasigroups \cite{Bat}.
We turn now to eq.\r{409}. We find then
\be
&&(\Om_a, \Om_b)_{\Om}=\{\Om_a, \{\Om, \Om_b\}\}=-\la_a^f\la_b^e
U_{ef}^d\pet_d(-1)^{\ve_d+\ve_e+\ve_f+\ve_b\ve_f}+\nn\\&&+
\left(\la^c_a\{f_c, \la^d_b \}-\la^c_b\{f_c,
\la^d_a\}(-1)^{\ve_a\ve_b}\right)\pet_d(-1)^{\ve_c}
\e{4271}
which implies that eq.\r{409}
here may  be written as
\be
&&\d_a\tilde{\la}_b^c-\d_b\tilde{\la}_a^c
(-1)^{\ve_a\ve_b}=\tilde{\la}^e_a
\tilde{\la}^d_b
\tilde{U}^c_{de}(-1)^{\ve_b\ve_e+\ve_c+\ve_d+\ve_e}.
\e{428}
where $\tilde{\la}_a^b\equiv V\la_a^b$ and
$\tilde{U}_{ab}^c \equiv V U_{ab}^c$, where
in turn the differential operator $V$
satisfies the equation
\be
&&\d_a V=V ad(H_a), \quad H_a\equiv \la_a^bf_b(-1)^{\ve_b}
\e{4281}
where in turn $ad(A)X=\{A,X\}$ for any
dynamical function $X$. The
integrability condition of \r{4281} is
\be
&&\d_aH_b-\d_bH_a(-1)^{\ve_a\ve_b}+\{H_a, H_b\}=0
\e{4282}
which is equivalent to \r{428}.

 One may note that
\be
&&Y_a(\phi)=\{\Om, \Om_a\}=\la^b_a
f_b(-1)^{\ve_a+\ve_b}+\nn\\&&+
\la^b_a\cC^dU^c_{db}\pet_c(-1)^{\ve_a+\ve_c}+\cC^b\{f_b,
\la^c_a\}\pet_c.
\e{429}
If we  assume $Q(\phi)$  to have zero Poisson
bracket with the last term in \r{429}, then
\r{403} reduces to
\r{302}, which was used for quasigroups in \cite{Quanti,OG}.

\subsection{Functions in arbitrary involutions in an antibracket sense.}
 Let
$\cE$ be an antisymplectic manifold and let
$f_a$ be functions on
$\cE$ satisfying the antibracket algebra
\be
&&(f_a, f_b)= U^c_{ab}f_c,\quad \ve(f_a)\equiv\ve_a,\quad
\ve(U^c_{ab})=\ve_a+\ve_b+\ve_c+1.
\e{430}
The structure coefficients $U^c_{ab}$
are arbitrary dynamical functions and
different from the ones in \r{401} due
to the oddness of the antibracket. This case is
most efficiently treated if we extend
the original antisymplectic manifold $\cE$ by
the ghost and antighost variables
$\cC^a,
\cC^*_a$,
$\ve(\cC^a)=\ve_a$, $\ve(\cC^*_a)=\ve_a+1$, satisfying
$(\cC^a,
\cC^*_b)=\del^a_b$. We may then always define an {\em even}  function $\Om$
with ghost number one according to the prescription
\be
&&\Om=\cC^a f_a+\ldots, \quad (\Om, \Om)=0,\nn\\
&&(G,\Om)=\Om,\quad G\equiv -\cC^a\cC^*_a
\e{431}
where $G$ is an {\em odd}  ghost number
function and where the additional terms in
$\Om$ depends on
$\cC^*_a$ (see
\cite{ASGT}).  In analogy with the previous
construction we define the generating
function
$R(\phi)$  by  the differential equation
\be
&&R(\phi)\stackrel{\lea}{\d_a}=
(R(\phi), Y_a(\phi)), \quad R(0)=R,
\e{432}
where $\phi^a$, $\ve(\phi^a)=\ve_a+1$
are parameters and where $Y_a(\phi)$,
$\ve(Y_a)=\ve_a$, satisfies the integrability conditions
\be
&&Y_a\stackrel{\lea}{\d_b}-Y_b\stackrel{\lea}{\d_a}
(-1)^{(\ve_a+1)(\ve_b+1)}=(Y_a,
Y_b),
\e{433}
which in turn are integrable. $Y_a(\phi)$ is of the form
\be
&&Y_a(\phi)=\la^b_a(\phi)f_b(-1)^{\ve_a+\ve_b}+
\{\mbox{\small possible ghost
dependent terms}\},\; \la^b_a(0)=\del^b_a,
\e{434}
which makes \r{432} similar to \r{310}. (However,
$\la^b_a(\phi)$ may be dynamical
functions here.) Since \r{432} implies
\be
&&(R(\phi), R(\phi))\stackrel{\lea}{\d_a}=
(Y_a(\phi), (R(\phi), R(\phi))),
\e{435}
the boundary condition $(R(0), R(0))=0$
 implies $(R(\phi), R(\phi))=0$ which
shows that $R\,\ra\, R(\phi)$ is an anticanonical transformation.

The generalized Poisson brackets are defined by
\r{214} with
$f_a$ replaced by
$Y_a$. In particular we have
\be
&&\{Y_a(\phi),
Y_b(\phi)\}'_{R(\phi)}\equiv R(\phi)\stackrel{\lea}{\d_a}
\stackrel{\lea}{\d_b}(-1)^{\ve_a}=\nn\\
&&=\half\left((Y_a,
(R(\phi), Y_b))-(Y_b, (R(\phi),
Y_a))(-1)^{\ve_a\ve_b}\right)+\nn\\
&&+\half(R(\phi),
Y_a\stackrel{\lea}{\d_b}+Y_b\stackrel{\lea}{\d_a}
(-1)^{(\ve_a+1)(\ve_b+1)})(-1)^{\ve_a}.
\e{436}
We
expect the last line to vanish at $\phi^a=0$
 in which case  \r{436} coincides with
\r{102}. Indeed, this is
the case  for the first rank quasigroup
theories treated below. For generalized
higher order Poisson brackets we should
 have a dual situation to the one in \r{407},
\ie we expect them to coincide with the
 generalized higher order Poisson brackets in
section 4.2 at $\phi^a=0$.
Although the above generalized Poisson
 brackets do not satisfy Leibniz' rule
according to \r{1022}, we have for the functions $Y_a$
\be
&&\{Y_aY_b, Y_c\}_R-Y_a\{Y_b, Y_c\}_R-\{Y_a,
Y_c\}_RY_b(-1)^{\ve_b\ve_c}=\nn\\ &&=
-\half
(R, Y_a)
 (Y_b, Y_c) (-1)^{\ve_b+\ve_c} -
 \half(Y_a, Y_c)(R, Y_b)(-1)^{\ve_c(\ve_b+1)},
\e{4361}
which vanishes at the hypersurface
determined by $\ca^*_a=0$ and $\theta_a=0$.

Analogously to \r{408} we propose here
the following form for the connections $Y_a$
in \r{432}
\be
&&Y_a(\phi)\equiv(\Om, \Om_a(\phi)),
\quad \ve(\Om_a)=\ve_a+1.
\e{437}
Comparison with \r{434}
requires then $\Om_a$ to be of the form
\be
&&\Om_a(\phi)=\la^b_a(\phi)\cC^*_b(-1)^{\ve_b}+
\{\mbox{\small possible ghost dependent
terms}\}.
\e{438}
The integrability conditions \r{433}
become in terms of $\Om_a$
\be
&&\Om_a\stackrel{\lea}{\d_b}-\Om_b\stackrel{\lea}{\d_a}
(-1)^{(\ve_a+1)(\ve_b+1)}=
\{\Om_a,
\Om_b\}_{\Om}-\half(\Om_{ab}, \Om),
\e{439}
where we have introduced the $\Om$-bracket
which is a Poisson bracket defined in
accordance with
\r{102},
\ie
\be
&&\{A, B\}_{\Om}\equiv\half\left((A, (\Om, B))-(B, (\Om,
A))(-1)^{\ve_A\ve_B}\right).
\e{440}
The form \r{438} means then that \r{439} is a new kind of
 generalized Maurer-Cartan equations for
$\la^b_a(\phi)$, $\ve(\la^b_a)=\ve_a+\ve_b$.
It is clear that the integrability
conditions of \r{439} will involve higher order Poisson brackets and
$\Om_{abc\cdots}$ with still more indices.

Also here we propose
one single master equation involving only
two basic dynamical functions which
determines
$\Om_a$,
$\Om_{ab}$, etc.  As in section 5.1 we first
make
$\phi^a$ dynamical but now by the
introduction of  antifields $\phi^*_a$. Then we
introduce new ghost variables $\eta^a$,
$\ve(\eta^a)=\ve_a$, which are to be
treated as parameters. By means of
$\phi^*_a$ and $\eta^a$ we define then the extended
{\em even} $\Om$-function, $\Delta$, defined by
\be
&&\Delta\equiv\Om-\eta^a \phi^*_a
(-1)^{\ve_a}, \quad (\Delta, \Delta)=0\nn\\
&&(\phi^a, \phi^*_b)=\del^a_b,\quad
\ve(\eta^a)=\ve(\phi^*_a)=\ve_a.
\e{441}
In addition we introduce an {\em odd}
function $S(\phi, \eta)$ defined by
\be
&&S(\phi, \eta)=G+\eta^a\Om_a(\phi)-\half
\eta^b\eta^a\Om_{ab}(\phi)(-1)^{\ve_b}+
\nn\\&&+{1\over6}\eta^c\eta^b\eta^a\Om_{abc}(\phi)
(-1)^{(\ve_a+\ve_b+\ve_c+\ve_a\ve_c)}
+\cdots\nn\\&&\cdots+
{1\over n!}\eta^{a_n}\cdots\eta^{a_1}\Om_{a_1\cdots
a_n}(\phi)(-1)^{(\ve_{a_1}+\ve_{a_2}+\ldots+\ve_{a_{n}}+
\ve_{a_1}\ve_{a_n}+n)}+\cdots,
\e{442}
where $G$ is the even ghost number function in \r{431}.
$\Om_{a_1\cdots a_n}(\phi)$
are dynamical functions with the properties
\be
&&\ve(\Om_{a_1\cdots a_n}(\phi))=
\ve_{a_1}+\cdots+\ve_{a_n}+1,\quad (G,
\Om_{a_1\cdots a_n})=-n\Om_{a_1\cdots a_n},
\e{443}
where the last relation implies that
$\Om_{a_1\cdots a_n}$ has ghost number
$-n$ . Thus, if we assign ghost
number one to $\eta^a$ then $\Delta$
has ghost number one and $S$ has ghost number
zero. We propose now that the functions
$\Om_{a_1\cdots a_n}(\phi)$ are
related to the ones in the integrability
conditions above and that they are determined
by the master equation (cf.\r{416})
\be
&&\{S, S\}_{\Delta}=(\Delta, S),
\e{444}
where the Poisson bracket is defined by \r{102}, \ie we have
\be
&&\{S, S\}_{\Delta}\equiv(( S, \Delta), S).
\e{445}
Consistency requires  $(\Delta, S)$ to satisfy
\be
&&(\Delta, \{S, S\}_{\Delta})=0\quad
\Lra\quad ((\Delta, S), (\Delta, S))=0.
\e{446}
The explicit form of $(S, \Delta)(=-(\Delta, S))$ is
\be
&&(S, \Delta)=\Om+\eta^a(\Om_a,
\Om)-\eta^b\eta^a\Om_a\stackrel{\lea}{\d_b}
(-1)^{\ve_b}-\half\eta^b\eta^a(\Om_{ab},
\Om)(-1)^{\ve_b}+\nn\\&&+\half\eta^c\eta^b\eta^a\Om_{ab}
\stackrel{\lea}{\d_c}(-1)^{\ve_b+\ve_c}+{1\over
6}\eta^c\eta^b\eta^a(\Om_{abc},
\Om)(-1)^{\ve_a+\ve_b+\ve_c+\ve_a\ve_c}+O(\eta^4).
\e{447}
Also here \r{444} is of the form $(G, \Om)=\Om$
with $G$ given by $S$ and $\Om$ by
$(S, \Delta)$ which shows that $S$
may be viewed as an extended ghost charge.

 To
zeroth and first order in
$\eta^a$ eq.\r{444} is satisfied identically. However,
to second order in $\eta^a$ it
yields
\r{439}. The consistency
condition \r{446} yields on the other hand exactly
\r{433} when \r{437} is used to
second order in
$\eta^a$.
 At the third order in $\eta^a$ the master equation \r{444} yields
\be
&&\Om_{bc}\stackrel{\lea}{\d_a}
(-1)^{\ve_b(\ve_a+1)}+\half\{\Om_a,
\Om_{bc}\}_{\Om}(-1)^{(\ve_a+1)(\ve_c+1)}+
cycle(a,b,c)=\nn\\&&=\{\Om_a,
\Om_b,
\Om_c\}_{\Om}(-1)^{(\ve_a+1)(\ve_c+1)}-
{2\over3}\{\Om'_{abc},\Om\},\nn\\
&&
\Om'_{abc}\equiv\Om_{abc}+{1\over8}\left((\Om_{ab},
\Om_c)(-1)^{(\ve_a+1)(\ve_c+1)}+cycle(a,b,c)\right),
\e{448}
where
the higher  $\Om$-bracket is defined in
accordance with \r{3131}. The above
equations are also integrability conditions of \r{439}. At higher orders in
$\eta^a$ the master equation
\r{444} yields equations involving still higher
$\Om$-brackets and functions
$\Om_{abc\ldots}$ with still more indices. We
expect these equations to coincide
 with the integrability conditions of \r{448}. This
chain of integrability conditions
will terminate as soon as a higher order Poisson
bracket is zero. Also here this
level should be connected to the rank of the theory,
\ie the highest power of $\cC^*_a$ in $\Om$.
 For instance, if $\Om$ is of rank
$r$  it would be natural to have $(\Om_{a_1}, \Om_{a_2},
\ldots, \Om_{a_n})_{\Om}=0$ for $n>r+1$
in which case $\Om_{a_1a_2a_3\cdots a_n}=0$
for $n>r$ (see below).

If we compare the  above formulas with
those of the previous subsection we notice
first that the master equations \r{416} and
\r{445} are exactly dual. Then we notice
that \r{439}, \r{441}, \r{442}, and \r{444}
 are dual to \r{409}, \r{413}, \r{414}, and
\r{419} with the simple prescription that
the Grassmann parities in the sign factors
are changed one step, $\ve\ra\ve+1$. The
same is true for \r{448} and \r{420} except
for an additional sign in the last line due
to the fact that we have made use of a
Jacobi identity which does not satisfy the simple rule above.

As an illustration of our formulas we
consider now functions $f_a$ forming a
quasigroup rank one theory in the
antibracket sense. In this case we have
(cf.\cite{ASGT})
\be
&&\Om=\cC^a f_a+\half \cC^b\cC^a
 U^c_{ab}\cC^*_c(-1)^{\ve_b}.
\e{449}
where $(\Om, \Om)=0$ requires
$(f_a, f_b)= U^c_{ab}f_c$ and
\be
&&U^d_{ab}U^e_{dc}(-1)^{\ve_b}-(f_a, U^e_{bc})
(-1)^{\ve_b}+{ cycle}(a,b,c)=0
\e{450}
plus conditions on $(U^c_{ab}, U^f_{de})$.
The latter conditions are satisfied if we
impose the conditions
\be
&&(U^c_{ab}, U^f_{de})=0, \quad ((f_d,
U^c_{ab}), U^g_{ef})=0,
\e{451}
which are stronger than what is
 required by $(\Om, \Om)=0$.
In this case $\Om_a$ may be chosen to be
\be
&&\Om_a(\phi)=\la^b_a(\phi)\cC^*_b(-1)^{\ve_b},
 \quad \la^b_a(0)=\del^b_a,
\e{452}
where
\be
&&(\la^b_a(\phi), \la^d_c(\phi))=0,
\quad \Lra \quad (\Om_a(\phi),
\Om_b(\phi))=0.
\e{453}
We require then
\be
&&(\Om_a, \Om_b, \Om_c)_{\Om}=0\quad \Lra \quad
(\{\Om_a, \Om_b\}_{\Om}, \Om_c)=0,
\e{454}
which  allows for the choice $\Om_{ab}=0$ since
\r{454} makes the Poisson bracket
$\{\Om_a, \Om_b\}_{\Om}$ satisfy the Jacobi identities
due to \r{209}, which in turn
makes
\r{439}  integrable without the need of a nontrivial $\Om_{ab}$. All higher
integrability conditions are then identically zero. Condition
\r{454} is satisfied if we impose the conditions
\be
&&(\la^b_a(\phi),
U^c_{de})=0,
\quad (\la^b_a,(\la^d_c,
f_e))=0.
\e{455}
All the above conditions specify  quasigroups in an antibracket sense.
We turn now to \r{439}. We find then
\be
&&\{\Om_a, \Om_b\}_{\Om}=(\Om_a,
(\Om, \Om_b))=-\la_a^f\la_b^e
U_{ef}^d\ca^*_d(-1)^{\ve_b+\ve_e+\ve_b\ve_f}-\nn\\&&-
\left(\la^c_a(f_c, \la^d_b )-\la^c_b(f_c,
\la^d_a)(-1)^{(\ve_a+1)(\ve_b+1)}
\right)\ca^*_d(-1)^{\ve_c+\ve_d}.
\e{4551}
Eq.\r{439} may then  be written as
\be
&&\d_a\tilde{\la}_b^c+\d_b\tilde{\la}_a^c
(-1)^{\ve_a\ve_b}=(-1)^{\ve_a}\tilde{\la}^e_a
\tilde{\la}^d_b
(-1)^{(\ve_b+1)(\ve_e+1)}\tilde{U}^c_{de}
(-1)^{\ve_c+\ve_d+\ve_e}.
\e{456}
where $\tilde{\la}_a^b\equiv W\la_a^b$ and
$\tilde{U}_{ab}^c \equiv W U_{ab}^c$, where
the differential operator $W$ satisfies
the equation
\be
&&\d_aW=W Ad(G_a), \quad G_a\equiv
\la_a^bf_b(-1)^{\ve_a+\ve_b}
\e{4561}
where in turn $Ad(A)X\equiv (A,X)$ for any function
$X$. The integrability condition
of \r{4561} is
\be
&&\d_aG_b-\d_bG_a(-1)^{(\ve_a+1)(\ve_b+1)}+(G_a, G_b)=0
\e{4562}
which is equivalent to \r{456}.

 One may note that
\be
&&Y_a(\phi)=(\Om, \Om_a)=\la^b_a
f_b(-1)^{\ve_a+\ve_b}+\nn\\&&+\la^b_a\cC^d
U^c_{db}\cC^*_c(-1)^{\ve_a}
+\cC^b(f_b,\la_a^c)\cC^*_c(-1)^{\ve_c}.
\e{457}
If we  assume $R(\phi)$  to have zero
antibracket with the last terms, then
\r{432} reduces to
\r{310}. It is amusing that in the quasigroup case
the expressions \r{449}, \r{452},
and \r{457} for $\Om$,
$\Om_a$, and $Y_a$ may be obtained from  \r{421},
\r{424}, and \r{429} by the
replacement $\pet_a\ra\ca^*_a(-1)^{\ve_a}$.

\setcounter{equation}{0}
\section{Conclusions.}
We have considered antibrackets on symplectic manifolds and Poisson brackets on
antisymplectic manifolds, as well as generating functions for these brackets. To
every relation in the first case we have found a dual one in the second case and
vice versa. In our presentation we have started with the simplest relations
which we
 have successively generalized to finally end up with the remarkable master
equations \r{416} and \r{444} representing generalized Maurer-Cartan
equations. It
might be useful to briefly summarize our results. If we consider a
$(2n,2n)$-dimensional manifold $\cE$ it may be either a symplectic or an
antisymplecic manifold. In the case $\cE$ is a symplectic manifold we define a
pre-antibracket or 2-antibracket by \r{101}. This bracket satisfies all
properties
of a conventional antibracket except for  the Jacobi
identities
\r{a12} and Leibniz' rule
\r{a13}. We claim that  if the odd function
$Q$ satisfies $\{Q, Q\}=0$ then we have a
consistent set of generalized antibrackets starting from the 2-antibracket
\r{101}. The higher order antibrackets may be expressed in
terms of the lower ones recursively. The general 3-antibracket is \eg given by
\r{3052}. Depending on the properties of
$Q$ there is a certain level at which the
higher order brackets  terminate. All these brackets are interrelated in
such a way
that the Jacobi identities for
\r{101} are not necessary for consistency.
It is remarkable that just these brackets enter in a natural way in the
master equation
\r{416}. In a way this demonstrates their importance. We have also given
generating
functions for these brackets: in section 3 for functions  satisfying Lie algebra
relations, and in section 4 for functions in arbitrary involutions in the
original
Poisson bracket. The  2-antibracket
\r{101} satisfies the Jacobi identities only if the 3-antibracket \r{3052}
vanishes
according to \r{3051}. This condition either
restricts $Q$ or the class of functions
to be considered. When the 3-antibracket vanishes then all higher antibrackets
vanish. Leibniz' rule
\r{a13} for the 2-antibracket leads then to further restrictions. In section 2 we
showed that if we restrict to a class of functions satisfying an abelian
algebra in
the Poisson bracket then
\r{101} is a true antibracket which can be non-degenerate on this submanifold of
$\cE$. Finally we have shown that all the above results are valid if $\cE$
instead is
an antisymplectic manifold on which we define the pre-Poisson bracket
\r{102}, \ie
all the above properties have their dual partners.  In this latter case we have
obtained a new type of generalized Poisson brackets involving higher
Poisson brackets
which are  different from the Nambu brackets \cite{Nambu}. The higher
brackets are directly connected to the violations of the Jacobi identities.
We have
shown that these generalized Poisson brackets appear naturally in a new type of
master equations for generalized Maurer-Cartan equations for open groups in an
antibracket sense.

Finally, we would like to  mention that we believe there  are also
dual equations to the classical counterparts of the new Sp(2)-antibrackets
and Sp(2)
master equations introduced in \cite{Sp2QA}.
\\
\\
\\
\noindent
{\bf Acknowledgments}

I.A.B. would like to thank Lars Brink for his very warm hospitality at the
Institute of Theoretical Physics, Chalmers and G\"oteborg University. The
work is partially supported by INTAS-RFBR grant 95-0829. The work of I.A.B.
is also
supported by INTAS grant 96-0308 and by RFBR grants 96-01-00482, 96-02-17314.
I.A.B. and R.M. are thankful to the Royal Swedish Academy of Sciences for
financial support.\\ \\

\def\theequation{\thesection.\arabic{equation}}
\setcounter{section}{1}
\setcounter{equation}{0}
\renewcommand{\thesection}{\Alph{section}}
\noindent
{\Large{\bf{Appendix A}}}\\ \\
{\bf Defining properties of the conventional Poisson bracket.}\\ \\
The
defining properties of the Poisson bracket, $\{f, g\}$,
for functions $f, g$ on a
manifold $\cS$ are \\
\\ 1) Grassmann parity
\be
&&\ve(\{f, g\})=\ve_f+\ve_g.
\e{a1}
2) Symmetry
\be
&&\{f, g\}=-\{g, f\}(-1)^{\ve_f\ve_g}.
\e{a2}
3) Linearity
\be
&&\{f+g, h\}=\{f, h\}+\{g, h\}, \quad (\ve_f=\ve_g).
\e{a3}
4) Jacobi identities
\be
&&\{f,\{g, h\}\}(-1)^{\ve_f\ve_h}+cycle(f,g,h)\equiv0.
\e{a4}
5) Leibniz' rule
\be
&&\{fg, h\}=f\{g, h\}+\{f, h\}g(-1)^{\ve_g\ve_h}.
\e{a5}
6) For any odd/even parameter $\la$ we have
\be
&&\{f, \la\}=0\quad {\rm any}\;f\in\cS.
\e{a6}
The Poisson bracket can only be nondegenerate if the dimension of the even
subspace of $\cS$ is even. ($\cS$ is then
a symplectic manifold.) If $z^A$ are
independent coordinates on $\cS$ then we have
explicitly ($\d_A\equiv \d/\d z^A$)
\be
&&\{f(z), g(z)\}=f(z)\stackrel{\lea}{\d}_A\omega^{AB}(z)\d_B\, g(z),
\e{a7}
where $\omega^{AB}(z)$ satisfies ($\ve_A\equiv\ve(z^A)$)
\be
&&\omega^{AB}(z)=-(-1)^{\ve_A\ve_B}\omega^{BA}(z),\quad
\ve(\omega^{AB}(z))=\ve_A+\ve_B,\nn\\&&
\omega^{AD}(z)\d_D\omega^{BC}(z)(-1)^{\ve_A\ve_C}+cycle(A,B,C)=0.
\e{a8}
The last equalities follow from the Jacobi identities \r{a4}.
\\
\\
\\
\setcounter{section}{2}
\setcounter{equation}{0}
    \noindent
    {\Large{\bf{Appendix B}}}\\ \\
 {\bf Defining properties of the conventional antibracket.}\\ \\
The defining properties of the antibracket $(f,g)$
for functions $f, g$ on a manifold  $\cA$ are\\
\\ 1) Grassmann parity
\be
&&\ve((f, g))=\ve_f+\ve_g+1.
\e{a9}
2) Symmetry
\be
&&(f, g)=-(g, f)(-1)^{(\ve_f+1)(\ve_g+1)}.
\e{a10}
3) Linearity
\be
&&(f+g, h)=(f, h)+(g, h), \quad (\ve_f=\ve_g).
\e{a11}
4) Jacobi identities
\be
&&(f,(g, h))(-1)^{(\ve_f+1)(\ve_h+1)}+cycle(f,g,h)\equiv0.
\e{a12}
5) Leibniz' rule
\be
&&(fg, h)=f(g, h)+(f, h)g(-1)^{\ve_g(\ve_h+1)}.
\e{a13}
6) For any odd/even parameter $\la$ we have
\be
&&(f, \la)=0\quad {\rm any}\;f\in\cA.
\e{a14}
The antibracket can only be nondegenerate if $\cA$ is a supermanifold with
the dimension $(n,n)$. ($\cA$ is then an
antisymplectic manifold.) If $z^A$ are
independent coordinates on $\cA$ then we have explicitly
\be
&&\{f(z), g(z)\}=f(z)\stackrel{\lea}{\d}_A E^{AB}(z)\d_B\, g(z),
\e{a15}
where $E^{AB}(z)$ satisfies
\be
&&E^{AB}(z)=-(-1)^{(\ve_A+1)(\ve_B+1)}E^{BA}(z),\quad
\ve(E^{AB}(z))=\ve_A+\ve_B+1,\nn\\&&
E^{AD}(z)\d_D E^{BC}(z)(-1)^{(\ve_A+1)(\ve_C+1)}+cycle(A,B,C)=0.
\e{a16}
The last equalities follow from the Jacobi identities \r{a12}.\\ \\

\end{document}